\documentclass[twocolumn,showpacs,preprintnumbers,amsmath,amssymb]{revtex4}
\usepackage{graphicx}
\usepackage{dcolumn}
\usepackage{bm}
\usepackage[latin1]{inputenc} 
\usepackage{amsmath}

\newcommand{\comment}[1]{}
\newcommand\etal{\mbox{\textit{et al.}~}}
\newcommand\eg{\mbox{\textit{e.g.}~}}
\newcommand\ie{\mbox{\textit{i.e.}~}}

\begin{document}
\setlength{\unitlength}{0.7\textwidth} \preprint{}

\title{Intrinsic rotation of toroidally confined  magnetohydrodynamics}

\author{Jorge A. Morales$^{1}$, Wouter J.T.  Bos$^{1}$, Kai Schneider$^{2}$, David C. Montgomery$^{3}$}

\affiliation{$^1$ LMFA - CNRS, Ecole Centrale de Lyon - Universit\'e de Lyon, Ecully, France}
\affiliation{$^2$ M2P2 - CNRS \& CMI, Aix-Marseille Universit\'e, Marseille, France}
\affiliation{$^3$ Department of Physics and Astronomy, Dartmouth College, Hanover, New Hampshire, USA}

\begin{abstract}
The spatiotemporal self-organization of viscoresistive magnetohydrodynamics (MHD) in a to\-roi\-dal geometry is studied. Curl-free toroidal magnetic and electric fields are imposed.  It is observed in our simulations that a flow is generated, which evolves from dominantly poloidal to toroidal when the Lundquist numbers are increased. It is shown that this toroidal organization of the flow is consistent with the tendency of the velocity field to align with the magnetic field. Up-down asymmetry of the geometry causes the generation of a non-zero toroidal angular momentum.
\end{abstract}


\maketitle

\paragraph{Introduction.} The magnetic confinement of fusion plasmas is strongly influenced by turbulent fluctuations. These fluctuations degrade the quality of the confinement and thereby reduce the performance of the fusion reactor. It was discovered three decades  ago \cite{Wagner1982}  that, under certain circumstances, the turbulent activity is reduced, leading to a better confinement. Still today the understanding of this low-to-high-confinement (LH) transition is far from complete. There is however strong evidence that large toroidal velocities of the plasma are a feature that is either at the origin, or a consequence of, the mechanism that is responsible for the LH-transition \cite{Rice1999,Rice2007}. Large toroidal velocities, of the order of several kilometers per second, are observed even in the absence of external momentum input. Several mechanisms are put forward to explain the toroidal rotation, mostly based on the turbulent transport of toroidal momentum generated at the tokamak edge (\eg in 
references \cite{Mattor1988,
Peeters2011,Diamond2009}). In this letter we present a mechanism which seems to be generic, since it is observed even in one of the coarsest descriptions of a fusion plasma: visco-resistive magnetohydrodynamics (MHD).

\begin{figure}
\includegraphics[width=1\linewidth]{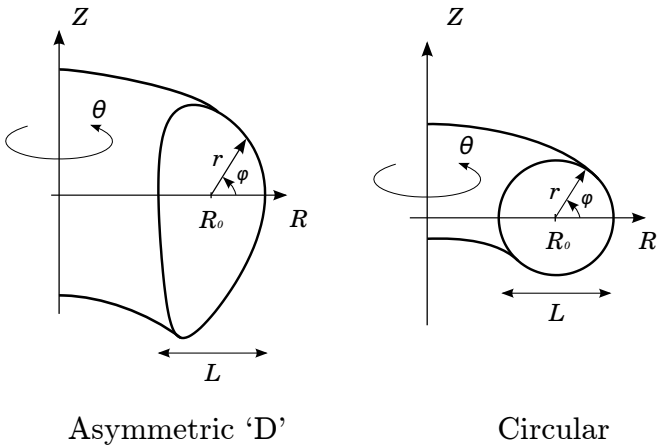}
\caption{Cross-sections of the toroidal geometries considered in the present work. The toroidal angle is labelled $\theta$ and the poloidal one $\varphi$. 
}\label{fig:tokamak_cross_section_shape}
\end{figure}

\paragraph{A MHD description of fusion plasmas.}
In the MHD description, the plasma is described as a charge-neutral conducting fluid. MHD, despite its low level of complexity compared to kinetic descriptions or two-fluid descriptions, gives already rise to a wealth of intricate phenomena and its analytical treatment is only possible in some simplified cases, either in the absence of velocity fields \cite{Grad1958,Shafranov1966} or in the absence of nonlinear interactions \cite{Bates1998}. We will come back to these analytical approaches, but before that, we present the equations that we consider. These are the dimensionless incompressible viscoresistive MHD equations for the velocity field $\bm u$ and for the magnetic field $\bm B$, in {\it Alfv\'enic} units \cite{Kamp2004},
\begin{eqnarray}
\frac{\partial \bm{u}}{\partial t}-M^{-1}\nabla^{2} \bm u&=&-\nabla\left(P+ \frac{1}{2}\,u^2 \right)+\bm u\times \bm \omega +\bm j\times \bm B,\label{eq:nondim1}\nonumber\\
\\
\frac{\partial \bm B}{\partial t}&=&-\nabla\times \bm E,~~~\\
\bm E&=&S^{-1} \bm j-\left[ \bm u\times\bm B\right],~~~\\
\nabla \cdot \bm u &=& 0, \hspace{1cm} \nabla \cdot \bm B = 0,~~~\label{eq:div}
\end{eqnarray}
with the current density $\bm j=\nabla \times \bm B$, the vorticity $\bm \omega=\nabla \times \bm u$, the pressure $P$ and the electric field $\bm E$.  These equations are non-dimensionalized using the toroidal Alfv\'en speed $C_A={B_0}/{\sqrt{\rho \mu_0}}$ as typical velocity, with $B_0$ the reference toroidal magnetic field at the center of the torus ($R=R_0$), $\rho$ the density and $\mu_0$ the magnetic constant. The reference length $L$ (see Fig.~\ref{fig:tokamak_cross_section_shape}) is the diameter of the cross section for the circular case and is the minor diameter for the asymmetric `D' shape ($L=1.88$ for both geometries). The dynamics are then governed by the initial and boundary conditions of the problem, and two dimensionless quantities: the viscous Lundquist number ($M$) and the Lundquist number ($S$) defined as  
\begin{equation}
M=\frac{C_A L}{\nu}, \hspace{1cm} S=\frac{C_A L}{\lambda},
\end{equation}
with $\lambda$ the magnetic diffusivity and $\nu$ the kinematic viscosity. The ratio of these two quantities is the magnetic Prandtl number $P_r={\nu}/{\lambda}$, which we have chosen unity in the present study, thereby reducing the number of free parameters, which characterize the magnetofluid, to one, the viscous Lundquist number. Previous investigations indicate that it is the geometric mean of the viscosity and the magnetic diffusivity which determines the dynamics \cite{Cappello2000,Shan1993-both}. In setting the Prandtl number to one, a change in the Lundquist numbers, $M$ or $S$, is equivalent to a change in the Hartmann number.

Let us now go back to the analytical description of visco-resistive MHD. In the static case in which $\bm u = 0$, equation (\ref{eq:nondim1}) reduces to an equilibrium 
\begin{equation}\label{eq:p=jb}
\nabla P=\bm j\times \bm B.
\end{equation}
In a cylindrical geometry this equilibrium can be achieved by various magnetic configurations such as the $z$-pinch or the $\theta$-pinch \cite{Freidberg1982}. In toroidal geometry it is problematic to obtain such an equilibrium, as we will now explain. We consider the case in which the driving toroidal electric field is curl-free within the plasma, over times of interest, such that $\bm E_\theta\sim 1/R$. Further do we assume the toroidal magnetic field to obey the same scaling, which follows from the integration of Amp\`ere's law on a toroidal loop. In the simplest case, we choose a space-uniform electrical conductivity such that the toroidal current induced by the electric field is also given by the same dependence, so that the externally imposed magnetic field  and toroidal, laminar, voltage-driven current density are given by,
\begin{equation}
\bm B_{0}(R) = B_0 \frac{R_0}{R} \bm e_{\theta}, 
 \hspace{1cm} \bm J_{0}(R) = J_0 \frac{R_0}{R} \bm e_{\theta}.
\label{eq:E0}
\end{equation}
Computing the Lorentz force resulting from these toroidal fields, taking into account the poloidal magnetic field induced by $\bm J_0$, results in a force-field which is not curl-free \cite{Montgomery1994}. Since the curl of the pressure gradient is necessarily zero, the equilibrium described by (\ref{eq:p=jb}) becomes impossible and additional terms of equation (\ref{eq:nondim1}) need to be taken into account to balance the equation. Since all other terms in (\ref{eq:nondim1}) are proportional to (or quadratic in) the velocity, the resulting state must be dynamic. That is, a toroidal plasma, described by visco-resistive MHD, confined by curl-free toroidal electric and magnetic fields, necessarily moves.  

It is true that the rationale described above depends on the choice of the electric conductivity, which was assumed to be uniform. It was however shown \cite{Bates1996,Montgomery1997} that to satisfy (\ref{eq:p=jb}) in a torus, very unusual profiles of the electrical conductivity must be assumed. We omit these rather unphysical cases and focus on the dynamical plasma behaviour which results for the simplest, uniform, conductivity profile. 

It follows from the foregoing that it is necessary to take into account all other terms in the MHD equations, and analytical treatment becomes impossible unless symmetries are assumed. To study the full dynamics we are obliged to solve numerically the system and this is what is done in the present investigation. Such fully three-dimensional non-stationary simulations, taking into account all relevant time and space-scales, are computationally demanding and only quite recently have the necessary resources and numerical methods become available to do such simulations.  Equations (\ref{eq:nondim1}-\ref{eq:div}) are discretized with a Fourier pseudo-spectral method on a Cartesian grid. To impose the boundary conditions we use the volume-penalization technique, a method of the immersed boundary type, which we consider a good compromise between the ease of implementation, flexibility in geometry, and the numerical cost of the simulation. Results for two dimensional MHD can be found in reference \cite{Bos2008-2}. 
We recently extended this method to study the three-dimensional visco-resistive MHD equations \cite{Morales2012}, and in the present 
communication we present the results of three-dimensional simulations in two toroidal geometries.

\begin{figure}
\includegraphics[width=0.8\linewidth]{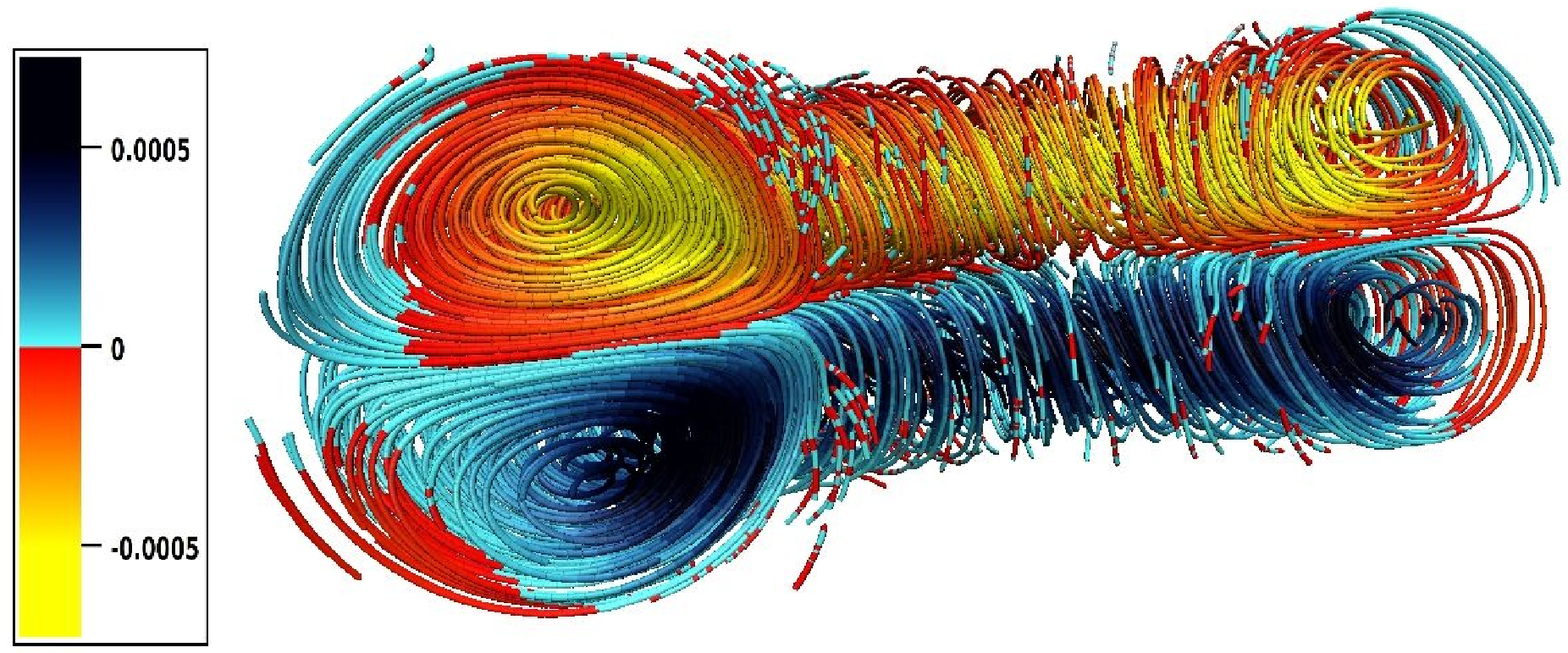}

\vspace{-1cm}

\includegraphics[width=0.8\linewidth]{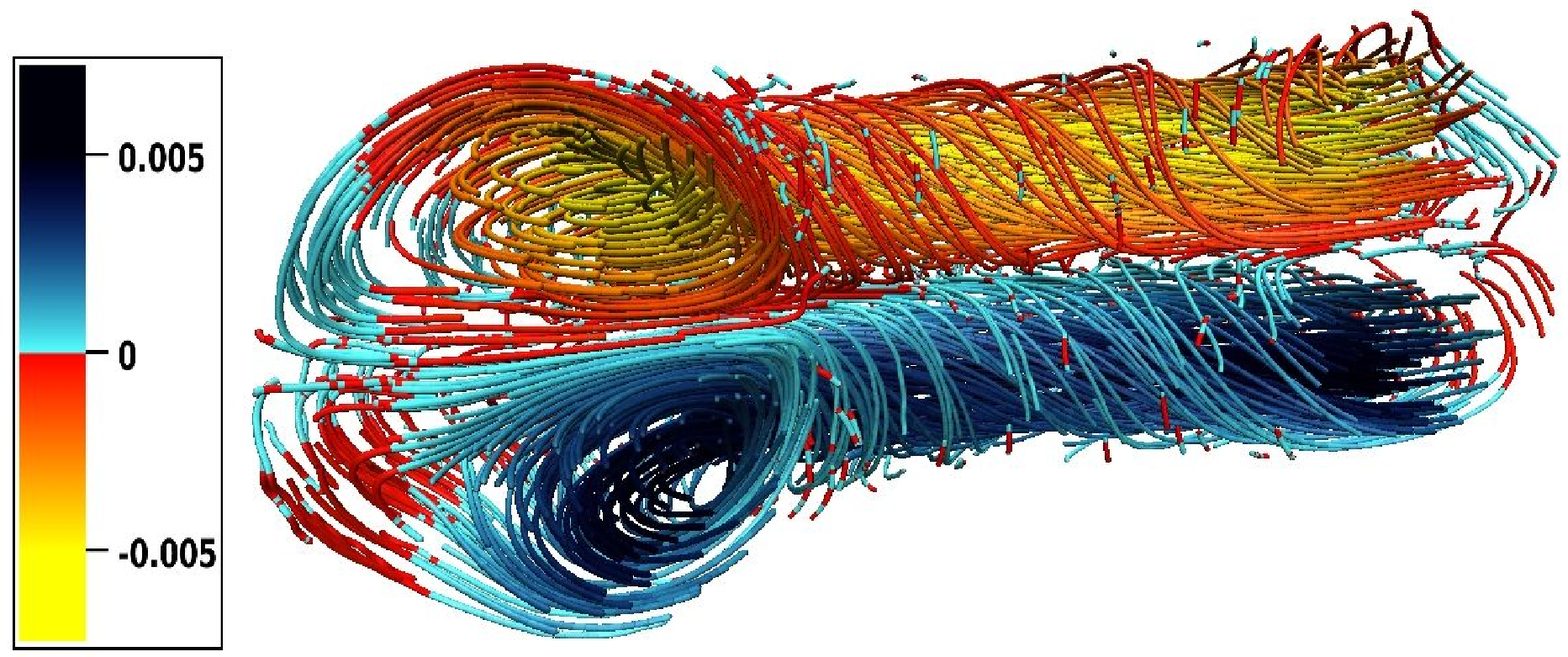}
\caption{Streamlines colored by the value of the toroidal velocity, $u_{\theta}$ for $M=15$ (top) and $M=150$ (bottom) in the geometry with circular cross-section. Only a part of the toroidal domain is shown.}
\label{fig:flow_ut_color_flow_Pr_1_Re_10_N256}
\end{figure}

\begin{figure}
\includegraphics[width=1.\linewidth]{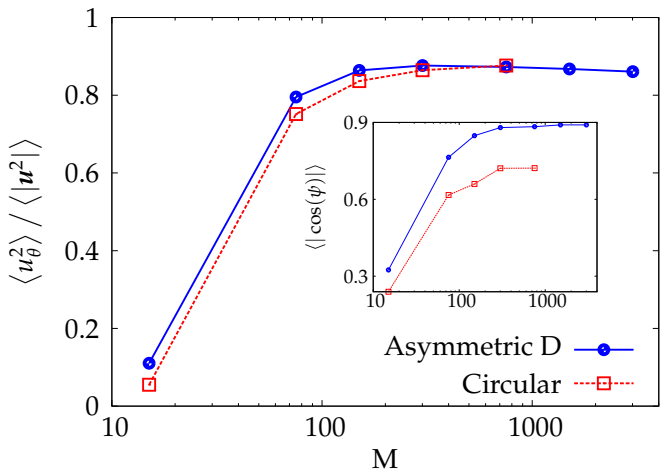}
\caption{The ratio of the mean-square toroidal velocity to the total mean-square velocity $\left< u_{\theta}^2\right> / \left<|\textit{\textbf{u}}^2|\right> $ as a function of  $M$. In the inset we show the average over the domain of the absolute value of the cosine of the angle between the velocity field and magnetic field.}
\label{fig:u_theta_rms_sur_u_rms_V02}
\end{figure}


\paragraph{Results of numerical simulations.}
Details of the numerical method are given in reference \cite{Morales2012}. Simulations are carried out on a cubic domain of size $2\pi$ consisting of $256^3$ grid-points for the highest values of $M$. The initial condition for the simulations is zero velocity, and no-slip velocity boundary conditions are imposed. We consider the boundaries of the fluid domain as perfectly conducting and coated with an infinitely thin layer of insulator. Thereby the normal component at the wall of the magnetic and current density fields vanishes. We impose toroidal magnetic and current density fields given by eq. (\ref{eq:E0}). the Biot-Savart law is used to determine the poloidal magnetic field induced by the toroidal current $\bm J_{0}(R)$.  All the simulations presented in this communication are performed with $B_0=0.8$ and $J_0=0.3$. This corresponds, for both geometries, to a pinch ratio $\Theta\approx0.16$, defined as the ratio between the wall-averaged poloidal and the volume-averaged toroidal magnetic 
field ($\Theta=\overline{B_{\varphi}}/\left<B_{\theta}\right>$). The only parameter that we vary is the  Lundquist number $M$. The simulations are time-dependent and they are stopped when a dynamical steady state is reached.


The results in  Fig.~\ref{fig:flow_ut_color_flow_Pr_1_Re_10_N256} show the presence of a poloidal flow, a pair of counter-rotating vortices in the $r-\varphi$ plane. For small $M$ the dynamics are dominantly poloidal, as is expected. Indeed, in the limit of vanishing nonlinearity, Bates \etal \cite{Bates1998} showed analytically that the steady state solution is a pair of poloidally rotating vortices, aligned with the toroidal direction. For non-zero nonlinearity, \ie, by increasing $M$, the vortices start moving in the toroidal direction, both in opposite direction. Their toroidal velocity increases with the Lundquist number $M$ in the two considered geometries. The three-dimensional velocity streamlines show a substantial change of topology from dominantly poloidal to dominantly toroidal flow (see Fig.~\ref{fig:flow_ut_color_flow_Pr_1_Re_10_N256}, bottom). This is quantified in Fig.~\ref{fig:u_theta_rms_sur_u_rms_V02}, where we observe that the principal direction of the flow motion is toroidal if  $M$ is 
raised beyond $\sim40$. The square toroidal velocity saturates for increasing $M$ at a value of $\sim86\%$ of the total square velocity. This toroidal organization of the flow is consistent with the tendency of the velocity-field to align with the magnetic field, as is illustrated in the inset of Figure \ref{fig:u_theta_rms_sur_u_rms_V02}, where we compute the average (over the toroidal domain) of the absolute value of the cosine of the angle between the velocity and magnetic field. This quantity is equal to one if the velocity and magnetic field are perfectly aligned or anti-aligned. It is shown that the trend towards a toroidal velocity follows exactly the same $M$ dependence as the alignment.

\begin{figure*}
a)\includegraphics[width=0.2\linewidth]{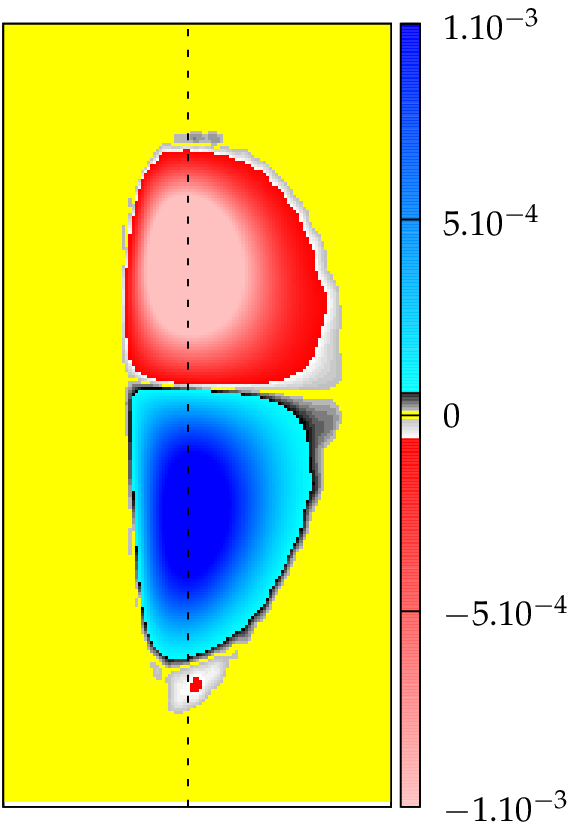}~
b)\includegraphics[width=0.2\linewidth]{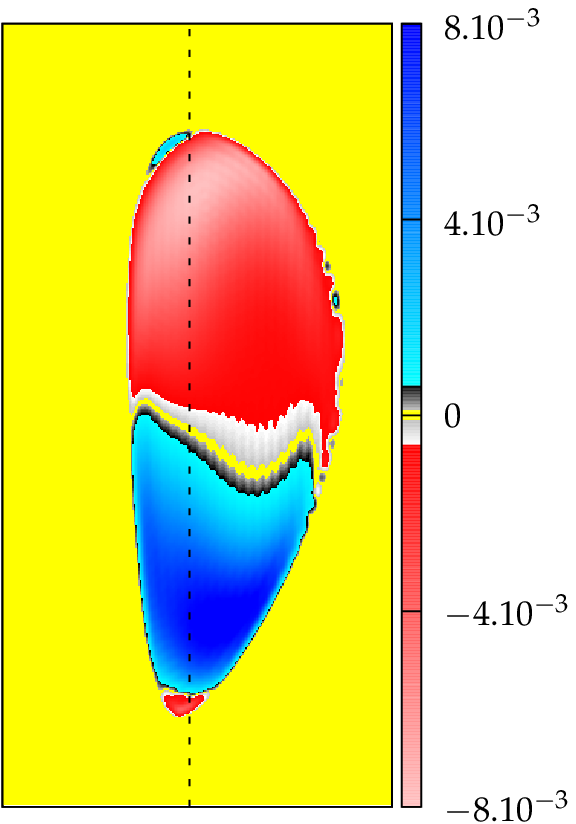}~
c)\includegraphics[width=0.2\linewidth]{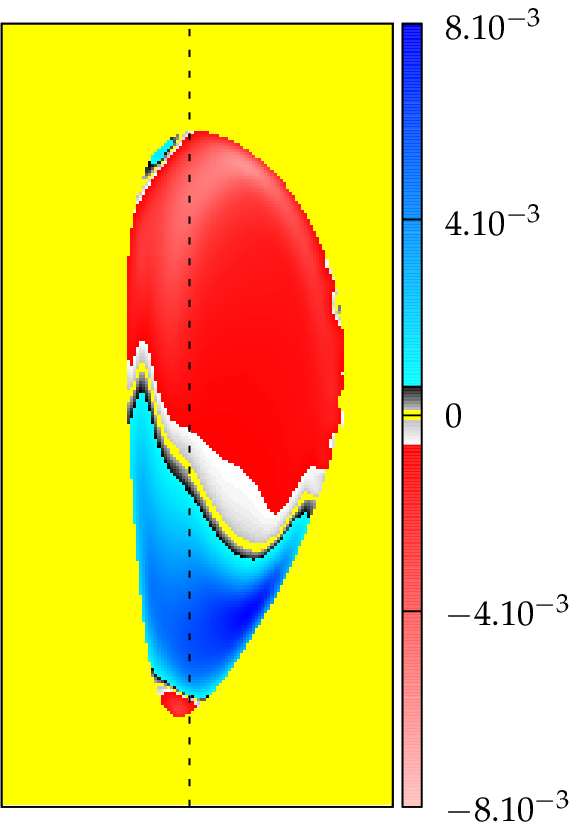}~
d)\includegraphics[width=0.25\linewidth]{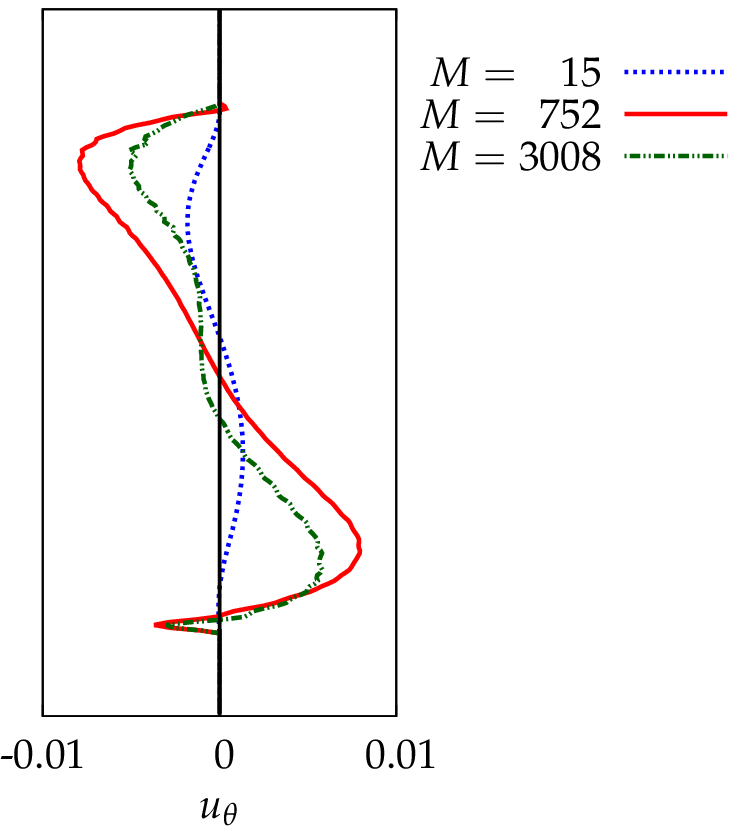}
\caption{Azimuthally averaged flow visualizations: toroidal velocity $u_{\theta}$ for $M=15$ (a), $M=752$ (b) and $M=3008$ (c).  (d) Toroidal velocity profiles along a vertical cut. The position of these cuts is indicated in (a), (b), (c) by a dotted vertical line. }
\label{fig:coupe_ut_color_asym_Pr_1_Re_10_N128_V02}
\end{figure*}

\begin{figure}
\includegraphics[width=.9\linewidth]{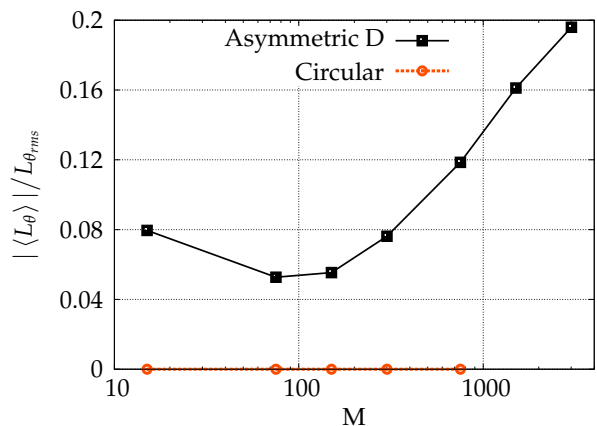}
\caption{Normalized toroidal angular momentum $\left|\left<L_{\theta}\right>\right|/L_{\theta_{rms}}$ as a function of  $M$ observed in the tori with asymmetric and symmetric cross-section, respectively.}
\label{fig:L_theta_mean_sur_L_theta_rms_V02}
\end{figure}

This tendency towards a dominant toroidal flow is similar for the torus with the asymmetric cross-section as is shown in Fig.~\ref{fig:u_theta_rms_sur_u_rms_V02}, while the alignment is even more pronounced. In both geometries, the generated velocity field contains non-negligible fluctuations. The quantities $\bm u'$ and $\bm B'$ denote the fluctuations around the azimuthally averaged instantaneous velocity and magnetic field, respectively. At $M=3008$ in the D-shaped geometry,  $\bm u'_{rms}/\bm u_{rms}=6.7\cdot 10^{-2}$, $\bm B'_{rms}/\bm B_{rms}=1.5\cdot 10^{-3}$ and $\bm u_{rms}/\bm B_{rms}=3.2\cdot 10^{-3}$. The rms values correspond here to volume averages over the toroidal domain. A detailed investigation of the spatial distribution of these fluctuations and its dependence on $M$ will be presented elsewhere.

A fundamental difference is observed between the flows that are generated in the two geometries. The volume-averaged toroidal angular momentum is defined by
\begin{equation}
\left<L_{\theta}\right>=\frac{1}{V}\int_{V} R  u_{\theta} dV.
\end{equation}
For the torus with circular cross-section, the time-average of this quantity is zero to a good computational approximation, due to the up-down symmetry of the observed flow. However, for the torus with asymmetric cross-section this is not the case. For low $M$, a poloidal pair of counter-rotating vortices appears (see Figure \ref{fig:coupe_ut_color_asym_Pr_1_Re_10_N128_V02} for azimuthally averaged flow visualizations, and toroidal velocity profiles along a vertical cut) as for the circular cross-section. Similarly, if the viscous Lundquist number is increased, an important toroidal flow develops. Unlike the symmetric case, there is a breaking of the symmetry in the flow and the part of the flow moving in the negative direction (the red zone), becomes larger at the expense of the part of the flow which moves in the positive toroidal direction (blue zone). This symmetry breaking, illustrated in Fig.~\ref{fig:coupe_ut_color_asym_Pr_1_Re_10_N128_V02}, leads to the development of a net toroidal flow. The 
toroidal 
angular momentum becomes hereby  non-zero (see Fig.~\ref{fig:L_theta_mean_sur_L_theta_rms_V02}). Its normalized value increases significantly with the viscous Lundquist number. The observed influence of up-down symmetry is consistent with axi-symmetric time-independent computations \cite{Kamp2004} and is also observed in gyrokinetic simulations and experiments \cite{Camenen2009,Camenen2010}. 

It is presently not clear if the velocity profile observed in our simulations will change qualitatively when $M$ is increased further and a transition to another flow-topology cannot be excluded. Also have we not yet investigated the influence of the magnetic Prandtl number. It is at this point perhaps important to say that we do not know what the viscosity should be to approximate the dynamics of experiments. However, the fact that this feature is observed in fully resolved simulations of the visco-resistive MHD equations is a result of major importance, since it shows how intrinsic toroidal rotation is present in one of the coarsest global descriptions of a fusion plasma, without invoking arguments on charge non-neutrality or kinetic theory.

\paragraph{Conclusion.}
We want to summarize the results that we obtained: considering curl-free toroidal electric and magnetic fields and constant transport coefficients, visco-resistive magnetofluids spontaneously generate velocity fields. This velocity field aligns (or anti-aligns) with the magnetic field, thereby generating a toroidal component. This is a nonlinear effect which becomes negligible in the limit of small Lundquist number. Furthermore, toroidal angular momentum is created, if the up-down symmetry of the torus is broken. 

 By its simplification, both in terms of the used model-equations, as well as in terms of the parameter range chosen for the properties of the conducting fluid, our investigation should be considered academic rather than directly applicable to the detailed description of existing machines. At the same time, since MHD does give a rough description of laboratory plasmas, the mechanism that we have described should be present, at least qualitatively, in existing devices. The observed magnetohydrodynamic self-organization thereby seems to be of major importance for the magnetically confined fusion community.

This work was supported by the contract \emph{SiCoMHD} (ANR-Blanc 2011-045), computing time was supplied by IDRIS, project 22206.


\end{document}